# Feasibility of Continuous Ventricular Volumetric Quantification in Arrhythmias using Real-Time 3D CMR-MOTUS

Thomas E. Olausson[1], Maarten L. Terpstra[1], Rizwan Ahmad[2], Edwin Versteeg[1], Casper Beijst[3], Yuchi Han[4], Marco Guglielmo[5,6], Birgitta K. Velthuis[5], Cornelis van den Berg[1], Alessandro Sbrizzi[1]

[1]Computational Imaging Group for MRI Therapy & Diagnostics, Center of Image Sciences, University Medical Center Utrecht, Utrecht, the Netherlands

[2]Department of Biomedical Engineering, The Ohio State University, Columbus, United States of America

[3]Department of Radiotherapy, University Medical Center Utrecht, Utrecht, the Netherlands

[4]Division of Cardiovascular Medicine, The Ohio State University, Columbus, Ohio, United States of America

[5]Department of Radiology, University Medical Center Utrecht, Utrecht University, Utrecht, the Netherlands

[6]Department of Cardiology, University Medical Center Utrecht, Utrecht University, Utrecht, the Netherlands

**Corresponding Author:** Thomas E. Olausson, Computational Imaging Group for MRI Therapy & Diagnostics, Center of Image Sciences, University Medical Center Utrecht, Utrecht, the Netherlands. Email: t.e.olausson@umcutrecht.nl

## Abbreviations

BMI: body mass index

bSSFP: balanced steady-state free precession

CMR: cardiovascular magnetic resonance

DIR: deformable image registration

EDV: end-diastolic volume

EF: ejection fraction

ESV: end-systolic volume

GRE: gradient echo

GT: ground truth

IRB: institutional review board

MRI: magnetic resonance imaging

OPRA: ordered pseudo radial

PVC: premature ventricular contraction

SGD: stochastic gradient descent

SNR: signal-to-noise ratio

SV: stroke volume

SVD: singular value decomposition

TV: total variation

# Key words

Free-Running; Real-Time; Motion fields; Arrhythmia; Premature ventricular contraction; Beat-to-beat variability; Volumetric quantification; Ejection fraction.

# Abstract

## Background

Conventional cardiovascular magnetic resonance (CMR) cine sequences rely on binning reconstructions that average multiple heartbeats, an assumption that breaks down in arrhythmic patients where beat-to-beat variations lead to motion artifacts and loss of clinically relevant functional information. While 2D real-time imaging can capture individual heartbeats, a stack of 2D slices is sub-optimal to map the full complexity of incoherent cardiac dynamics during arrhythmia. We investigated the feasibility of 3D real-time motion field reconstruction for continuous beat-to-beat volumetric quantification in patients with premature ventricular contractions (PVC) using a free-running CMR protocol.

## Methods

We extended CMR-MOTUS to jointly reconstruct real-time 3D motion fields and a motion-corrected reference image from continuously acquired data without breath-holds or ECG gating. A variable-density Cartesian sampling trajectory (OPRA) was used with a 3D spoiled gradient echo or balanced steady-state free precession sequence. The real-time volumetric beat-to-beat changes were quantified by propagating a single manual segmentation on the reference image, through all time frames using the reconstructed motion fields. The method was evaluated on a cardiac motion phantom with ground-truth static acquisitions and in 10 healthy volunteers and 10 patients with PVC. All in vivo datasets were acquired post contrast enhancement. The ejection fraction (EF) was compared to ground-truth values for the phantom and to standard 2D real-time cine EF measurement techniques for the in-vivo subjects.

### Results

Reconstructed mean EF values of the phantom experiment were close to the ground-truth (Mean EF = 17.86 ± 0.33% versus 17.27%). In healthy volunteers, narrow beat-to-beat EF distributions reflected normal physiological consistency. In PVC patients, the method revealed broader, sometimes bimodal EF distributions. Simultaneously acquired ECG signals supported the temporal correspondence between volume irregularities and PVC episodes.

### Conclusions

3D real-time joint motion field and image reconstruction from a free-running CMR protocol enables continuous beat-to-beat volumetric quantification in arrhythmic patients, revealing functional heterogeneity that conventional single-beat and averaging measurements (binning and gating) can obscure. Larger studies are needed before claims of clinical validation or replacement of standard 2D cine analysis.

## Background

Cardiovascular MRI (CMR) is the gold standard for non-invasive volumetric assessment of cardiac structure and function.[1] Traditional volumetric indicators such as ejection fraction (EF), end-diastolic volume (EDV) and stroke volume (SV) are widely used to quantify cardiovascular diseases.[2] However, these global parameters lack regional information and therefore fail to detect regional abnormalities despite normal EF, EDV and SV values.[2], [3], [4] A study has demonstrated the added clinical value in quantifying subtle regional motion variations using deformable image registration (DIR) to monitor progression and responses to treatment of cardiovascular diseases.[2] Other studies show that DIR can provide strain, strain rate, and velocity measurements while enabling automated computation of the standard volumetric quantities (i.e., EF, EDV,

SV) through a one-stop CMR post-analysis workflow.[5] More recently, automated motion-based functional assessment methods using image registration and deep learning have further shown that cine MRI can be used to derive ventricular function and cardiac mechanics in a largely operator-independent manner [6], [7],[8].

In order to accurately describe regional function, CMR cine images need to be of high quality with respect to spatiotemporal resolution and SNR. Due to the inherently low sampling rate of MRI, standard CMR cine sequences rely on binning reconstructions that average multiple heartbeats during the acquisition to achieve acceptable image quality.[9] Recent approaches have improved the coverage and motion robustness of these sequences[10], [11], [12], [13], as well as shown that single-breath-hold 3D cine imaging is feasible.[14], [15] While various motion surrogate signals and regularization strategies have been developed to improve binning methods, they fundamentally assume near identical cardiac cycles throughout the acquisition.[16], [17], [18] This assumption breaks down for arrhythmic patients, where beat-to-beat variations in morphology and function lead to motion artifacts and degrading image quality.[9], [19] Some approaches discard data acquired during arrhythmic episodes (arrhythmia rejection), but this reduces the acquisition efficiency when arrhythmia is frequent and loses valuable functional information during these arrhythmic events.[20] Additionally, respiratory motion can influence the hemodynamics and morphology of the heart.[21] Currently the clinical CMR methods for arrhythmia rely on 2D real-time MRI to capture these changes during breathing.[22] These methods ensure volumetric coverage by acquiring a stack of 2D images in rapid succession. However, using a stack of 2D imaging is sub-optimal to map the complexity and incoherence of heart dynamics, as the arrhythmic beats are captured on some slices but not others. Hence, 3D real-time imaging may enable more coherent assessment of cardiac hemodynamics and morphology, including the effects of arrhythmias.

Real-time MRI offers an alternative to binning methods by continuously acquiring data and reconstructing each heartbeat separately, visualizing the true physiological cardiac morphology without cardiac-cycle averaging.[19] In this paper, real-time refers to retrospective offline reconstruction of individual time frames from continuously acquired ungated data without cardiac-cycle binning. The high spatiotemporal requirements to measure real-time cardiac dynamics forces high undersampling of the MRI acquisition ($R>10$) which makes the reconstruction problem very challenging. Several methods have proposed to address the extreme undersampling, using innovative acquisition and reconstruction techniques. However, many rely on challenging non-Cartesian k-space sampling trajectories that can produce eddy current artifacts and require lengthy reconstruction times due to repeated gridding operations.[23], [24]

Recently, the ML-DIP framework successfully used a temporally incoherent variable density Cartesian sampling for 3D real-time reconstruction at a high nominal frame rate, capturing beat-to-beat variability in arrhythmic patients. However, the reconstructed image components in ML-DIP are not guaranteed to be motion-static, meaning the motion fields may not fully capture the motion dynamics, limiting their clinical interpretation[25]. Separately, CMR-MOTUS demonstrated a framework for jointly reconstructing real-time cardiorespiratory motion fields and contrast-varying motion-static images using low-rank assumptions, achieving promising 2D results with both Cartesian and non-Cartesian sampling.[26] This disentanglement of physiological motion in CMR-MOTUS with motion fields could enable a comprehensive cardiac functional assessment and the continuous beat-to-beat volumetric analysis workflow in arrhythmic patients.

This study represents an early technical feasibility study of real-time 3D imaging and motion field characterization. To achieve this, we extend CMR-MOTUS to 3D using an efficient Cartesian sampling pattern to reconstruct real-time cardiovascular motion fields at 20 Hz. Our approach leverages explicit, interpretable regularization strategies rather than an implicit deep learning regularization, and employing a low-rank motion field model for the reduction of the number of optimizable parameters. The co-reconstruction of 3D+t motion fields inherent to CMR-MOTUS simplifies downstream tasks such as real-time segmentation propagation for volumetric quantification in arrhythmic patients who would otherwise be ineligible for standard analysis. Due to the absence of in-vivo 3D+t ground truth (GT) data, we evaluate the reconstructed motion fields and volume quantification using a cardiac insert in a mechanical motion phantom[18]. In addition, we analyse data from 10 healthy volunteers and 10 arrhythmic patients and compare our extracted volumetric quantification parameters to the values derived from 2D imaging using clinically accepted software.

# Methods

## Framework

CMR-MOTUS jointly solves for a motion-corrected reference image and real-time deformation vector fields by solving the following minimization problems

$$\min_{\mathbf{q}} \sum_{t}^{M} \| \mathbf{F}( \mathbf{q} \mid \mathbf{D}_t ) - \mathbf{s}_t \|_2^2 \qquad (1)$$

$$\min_{[\mathbf{D}_1,\ldots,\mathbf{D}_M]=\boldsymbol{\Phi}\boldsymbol{\Psi}^{\mathrm{T}}} \sum_{t}^{M} \| \mathbf{F}(\mathbf{D}_t \mid \mathbf{q}) - \mathbf{s}_t \|_1 + \lambda \|TV(D)\|_{2,1} \tag{2}$$

where $q$ represents the motion-corrected reference image, $D_t$ are the real-time motion fields at each of $M$ time points, $s_t$ is the acquired k-space data at time $t$, and $F$ denotes the forward encoding operator that includes coil sensitivity multiplication, Fourier transform, k-space sampling mask, and the backwards warping operation.

The motion fields are parameterized as dense displacement vectors, where each voxel is assigned a 3D displacement vector. To reduce memory requirements, we use an explicit low-rank factorization $D = \Phi\Psi^T$, where $\Phi \in \mathbb{R}^{Nd\times R}$ contains $R$ spatial basis components for $N$ voxels across $d = 3$ spatial dimensions, and $\Psi \in \mathbb{R}^{M\times R}$ contains the corresponding temporal basis components. This explicit factorization avoids computing singular value decompositions (SVD) during optimization in the implicit factorization case, which would be computationally expensive for 3D data at 20 Hz temporal resolution. The full motion field $D \in \mathbb{R}^{Nd\times M}$ is recovered through matrix multiplication $D = \Phi\Psi^T$, substantially reducing the number of parameters compared to directly optimizing all displacement vectors across time. In the current implementation, $\Phi$ and $\Psi$ are optimized jointly through the shared data-consistency in equation 2 and motion regularization terms acting directly on the product $D$, rather than being estimated independently with separate penalties.

An $\ell_1$ norm is used in the data consistency term as motivated in [27] to reduce the influence of inflow effects cause by blood entering the imaging volume, which should not be captured by the motion fields. The regularization term $TV(D)$ applies isotropic total variation (TV) in the spatial domain to promote spatially smooth motion fields while preserving sliding motion with the mixed $\ell_2$-$\ell_1$ norm as described in [28]. The specific value of $\lambda$ was empirically selected based on visual assessment of the smoothness of the motion fields for each dataset. Automatic $\lambda$ selection remains future work.

For the following acquired datasets, we do not expect dynamic contrast enhancement. Hence, we adjusted the CMR-MOTUS image reconstruction step to only reconstruct a time-independent reference image, instead of a time-dependent reference image. This modification distinguishes our 3D implementation from the original 2D CMR-MOTUS framework[26], which used a low-rank plus sparse decomposition to model temporal contrast variations. While that decomposition effectively captured inflow effects in 2D, extending it to 3D at 20 Hz would require computing singular value decompositions on matrices with thousands of frames at each iteration, making the approach computationally expensive. We therefore represent the static-reference formulation as a practical approximation rather than as demonstrated robustness to inflow. Importantly, a time-independent reference image cannot fully represent inflow, temporal contrast variations, phase decoherence, or other non-deformation effects.

## Data Acquisition

### K-Space Sampling Trajectory

All experiments used the Ordered Pseudo RAdial (OPRA) k-space sampling trajectory[29]. For this Cartesian sampling trajectory, we acquired readout lines along the FH direction. In the axial plane, readouts are sequentially acquired along an “L”-shaped leaflet which rotates in the plane using the golden angle. At lower spatial frequencies, spacing between samples is reduced compared to higher spatial frequencies resulting in a variable density in the axial plane. The spacing is controlled by two user-specified parameters, namely the number of readouts per “L”-shaped leaflet and the density of lower spatial frequencies sampled in the axial plane. For these settings, and all other settings, we used the default settings as described in[29], with an exception to the k-space matrix size.

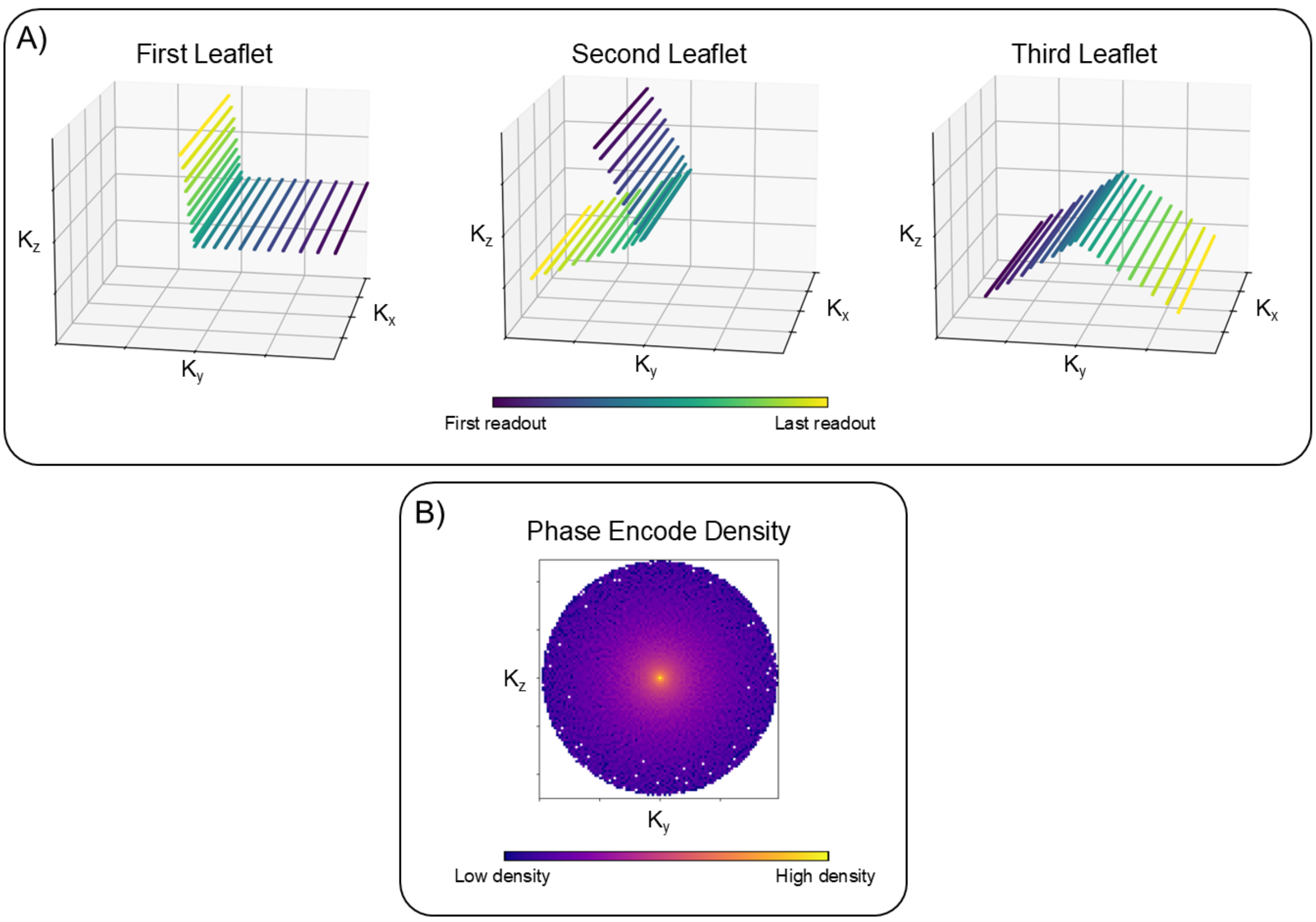

*Figure 1: Illustration of the Ordered Pseudo RAdial (OPRA) k-space sampling trajectory using default settings as described in [29]. Readouts are acquired along the kx axis. (A) Shows the k-space samples of the first three frames individually in each subplot. Each subplot contains one of the leaflets at separate golden angle increments. Note the small consecutive jumps between each readout which should minimize the effects of eddy currents. (B) Shows the cumulative sum of total number of readouts in the entire acquisition. Note the variable density in the ky-kz plane where lower spatial frequencies are sampled more often than higher spatial frequencies.*

## Phantom Study

We assessed the proposed approach using a cardiac insert[30] for the Modus QA motion phantom (IBA Dosimetry GmbH, Schwarzenbruck, Germany). In short, the deformable phantom is driven by a piston which compresses the liquid into two expandable silicone containers, which are shaped to mimic the left and right ventricles of a human heart. Figure 2A shows the experimental setup on the MRI table with the phantom. The silicon material and liquid have similar relaxation properties to myocardium and blood respectively (Left ventricle: T1 = 763 ± 76 ms, T2 = 49 ± 12 ms. Right ventricle: T1 = 775 ± 69 ms, T2 = 185 ± 35 ms. Blood: T1 = 1130 ± 44 ms, T2 = 155 ± 16 ms.) For the maximum piston motion amplitude achievable (and EF) with this phantom design, we operated the motion phantom at 60 oscillations per minute using a simple sinusoidal function.

A)

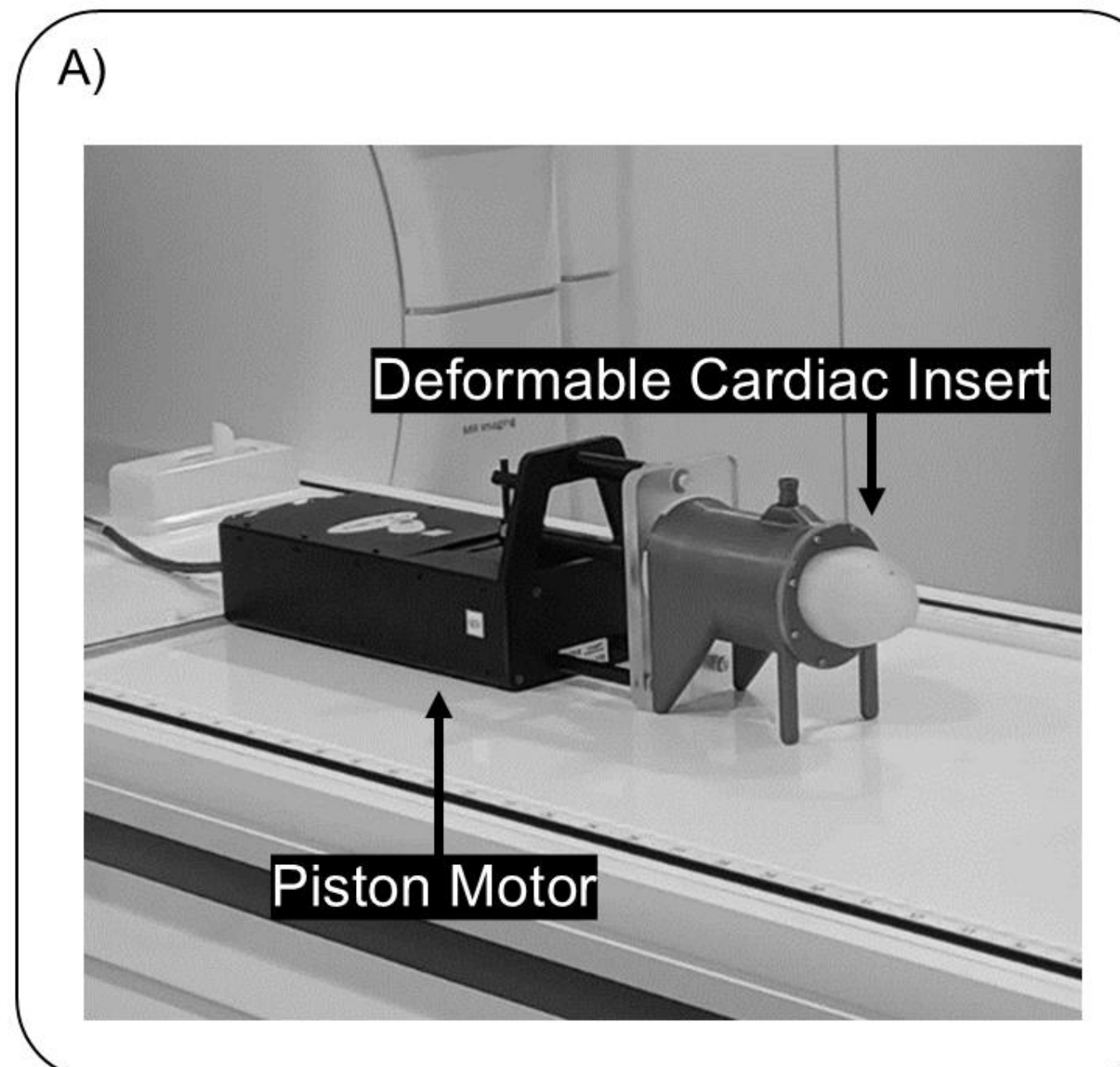

B)

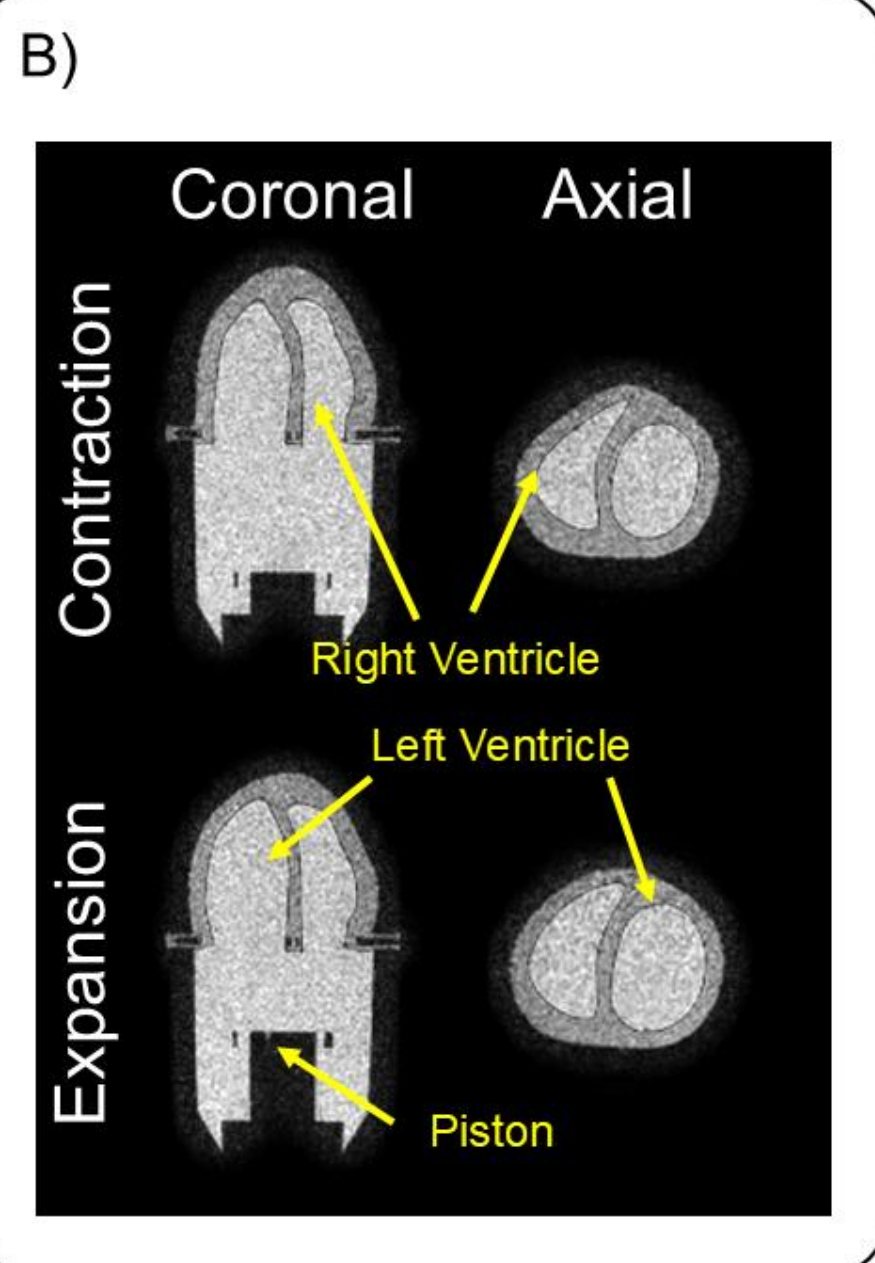

*Figure 2: Illustration of phantom anatomy. (A) shows the experimental setup of the cardiac phantom on the MRI table. The phantom is a silicone based deformable insert with two compartments to mimic the right and left ventricle of a human heart. Liquid inside the compartment is pushed by a piston which makes the phantom expand and contract. For our experiment we used the left ventricle for the analysis. (B) shows the 3D reference images acquired when the piston is static at largest displacement (expansion) and lowest displacement (contraction). These images were later used to segment the left ventricle and determine the ground truth ejection fraction. The yellow arrows label the anatomy of the phantom.*

All phantom acquisitions were performed on a 1.5T MRI scanner (Philips Ingenia, Philips Healthcare, Best, The Netherlands). We used a 3D T1-weighted spoiled gradient echo sequence with the following parameters: TR/TE = 2.4/1.07 ms, flip angle = 6°, field of view = 117 × 173 × 120 $mm^3$, spatial resolution = 2.00 × 2.00 × 2.00 $mm^3$. Total scan duration was approximately 96 seconds corresponding to 40,000 readouts. This duration was chosen empirically because shorter time-averaged reconstructions (inverse Fourier transform of all kspace points pooled) were too noisy for stable reference image initializations and motion estimation. For ground-truth validation, we acquired two static 3D images with the motionless piston at its peak positions (maximum and minimum displacement, respectively). These reference acquisitions used the same sequence parameters as the dynamic acquisitions but employed fully sampled linear Cartesian k-space sampling. Figure 2B shows images from these reference acquisitions with the inner anatomy of the phantom labelled.

## In Vivo Study

All in vivo studies were approved by the institutional review board, and written informed consent was obtained from all participants. We evaluated the method in vivo with two subject groups: healthy volunteers and patients with premature ventricular contraction (PVC). The PVC group consisted of 6 females and 4 male (age 48.6 ± 16.4 years, BMI 24.3 ± 4.5 kg/m$^2$), while the control group consisted of 6 females and 4 males (age 31.4 ± 9.7 years, BMI 27.2 ± 5.3 kg/m$^2$). Imaging was performed on two 1.5 and 3 Tesla MRI systems (MAGNETOM Sola and MAGNETOM Vida, Siemens Healthineers, Erlangen, Germany). PVC patients were scanned on the Sola system using a 3D balanced steady-state free precession (bSSFP) sequence approximately 20 minutes after the administration of a gadolinium-based contrast agent. Healthy volunteers were scanned on the Vida system using a 3D gradient echo sequence post ferumoxytol contrast enhancement. Sequence-specific imaging parameters for each group are summarized in Table 1.All acquisitions were performed using a free-running protocol, with free-breathing and no ECG gating. The total scan time for each acquisition was approximately 2 minutes corresponding to 40,000 readouts, as with the phantom study, this was empirically determined by visually assessing the time-averaged reconstruction. Shorter scans may be feasible, but were not systematically optimized in this study.

For contextual clinical comparison, free-breathing 2D real-time cine short-axis stacks covering the left ventricle from base to apex were acquired sequentially and analyzed using dedicated cardiac imaging software (SuiteHEART, NeoSoft, Pewaukee, Wisconsin) according to SCMR post-processing guideline [31]. In healthy volunteers, the 2D sequence was an undersampled spoiled gradient-echo research acquisition with variable-density golden-ratio sampling, temporal resolution of approximately 37-39 ms, spatial resolution of approximately 2.0-2.1 mm, and an acquisition time of about 6 seconds per slice. In PVC patients, 2D real-time imaging was performed with a balanced stead-state free precession sequence at similar spatial and temporal resolution and about 3 seconds per slice. For volumetric analysis, the 2D-derived EF was based on the first dominant sinus contraction identified during the slice-wise analysis. Because the stack was acquired slice by slice over multiple heartbeats rather than from the same heartbeat across the full stack, the derived 2D EF is treated here as a clinical comparator rather than as a beat-matched reference standard.

*Table 1: Summary of MRI acquisition parameters for phantom validation and in vivo studies.*

| Imaging Parameters | Phantom | Healthy Volunteers | Patients |
| --- | --- | --- | --- |
| Scanner | Philips Ingenia, 1.5 T | Siemens MAGNETOM Vida, 3 T | Siemens MAGNETOM Sola, 1.5 T |
| Sequence | 3D T1w spoiled GRE | 3D GRE with ferumoxytol | 3D bSSFP |

| Flip angle (°) | 6 | [14-18] | [32-40] |
|---|---|---|---|
| TR/TE (ms) | 2.4/1.07 | [3.10-3.17]/[1.22-1.24] | [2.89-3.03]/[1.28-1.34] |
| Field of view ($mm^3$) | 117 × 173 × 120 | [180-213] × [240-270] × 160 | [150-200] × [225-262] × [160-208] |
| Spatial resolution ($mm^3$) | 2.00 × 2.00 × 2.00 | [1.52-1.88] × [1.67-1.88] × 2.00 | [1.53-1.75] × [1.56-1.82] × 2.00 |
| Scan duration (s) | ~96 | ~120 | ~120 |

## Implementation

The following implementation details were used in all experiments in this study. Algorithmic settings were empirically optimized based on visual assessment of motion field smoothness and stability. Specific values of $\lambda$ ranged from $10^{-7}$ to $10^{-9}$ in magnitude per dataset. The reconstruction framework was implemented in Python using PyTorch for GPU-accelerated optimization and automatic differentiation. Coil sensitivity maps were estimated from the time-averaged k-space data using the ESPiRIT[32] algorithm. The motion-corrected reference image $q$ was initialized with the time-averaged reconstruction of the acquired k-space data. Motion fields were initialized with spatial low-rank components set to zero and temporal components initialized as smooth oscillating functions with random frequencies between 0.5-1 Hz.

The alternating minimization was performed for a maximum of 20 iterations, which was observed to be typically required to reach a plateau in the cost function. *To stabilize the early updates of the alternating optimization, $\Phi$ was spatially smoothed with a Gaussian kernel whose sigma was reduced linearly from 2.0 to 0 across these 15 alternating iterations. The final 5 iterations were then performed without any spatial smoothening.* During optimization of the motion fields parameters (equation 2), the temporal components $\Psi$ were passed through a Butterworth lowpass filter with a smooth cutoff at 4 Hz to enforce temporal smoothness and remove high-frequency noise. Optimization was performed using stochastic gradient descent (SGD) as implemented in PyTorch. For each backpropagation step, 20 temporal frames were randomly sampled to compute the gradient. This batch size was selected empirically to minimize noise in the reconstructed motion fields while maintaining reasonable convergence speed. For all experiments, hyperparameters were empirically tuned for best visual results in the motion warped images.

A total of 40,000 k-space readouts were acquired and reconstructed into real-time frames by grouping 20 consecutive readouts per frame, resulting in 2,000 frames spanning the entire acquisition. This corresponds to a nominal reconstructed frame rate of roughly 16-20 Hz depending on TR. However, the temporal coefficients are low-pass filtered, so the effective temporal information bandwidth is lower than the nominal frame spacing and should be interpreted separately from the sampling rate. This means that the current implementation is suited for continuous beat-to-beat volumetric quantification and the timing of the main contraction events, but not for resolving finer high-frequency details of PVC morphology. These faster components may be smoothed by the 4 Hz temporal model. The approximately 2-minute scan duration was chosen pragmatically to balance volumetric coverage, SNR, and the likelihood of capturing multiple respiratory cycles and arrhythmic events. It was not derived as a formal optimum.

Reconstructions were performed on a computational server equipped with 2 NVIDIA L40S GPUs, with typical reconstruction time of approximately 5 minutes per dataset.

## Data Analysis

To establish GT phantom data, manual segmentations of the phantom's left ventricle were performed in each of the two motion-static 3D images acquired at the piston's peak displacement positions using *3D Slicer*. We selected the left ventricle in the phantom data for the analysis. The maximum volume segmentation defined EDV and the minimum volume defined end-systolic volume (ESV), from which GT EF was calculated as

$$EF = \frac{(EDV - ESV)}{EDV} \times 100\%.$$

For both phantom and in vivo dynamic CMR-MOTUS datasets, a single manual segmentation was performed on the reconstructed motion-corrected reference image $q$ using the same method. As with the GT phantom data, the left ventricle was segmented using the 3D motion-corrected image $q$. For in vivo data, the left ventricle blood pool was segmented in the 3D motion-corrected image $q$ using *3D slicer*. The segmentation mask was propagated through all reconstructed time frames using the real-time motion fields $D$, enabling beat-to-beat volumetric quantification throughout the acquisition.

Volume curves were extracted from the real-time propagated segmentations, and peaks (EDV) and troughs (ESV) were detected using the *find_peaks* function from the *SciPy* library in Python. For each detected peak, the algorithm identified the subsequent trough to calculate EF for that specific cardiac contraction. Mean EF and its standard deviation across all detected cycles within each dataset were computed. For cases where a bimodal distribution was derived, the dominant mode was used for determining the mean EF. The primary validation of motion field accuracy was performed by comparing these real-time EF estimates to the GT measurements from the motion-static acquisitions of the same phantom as well as the mean Dice and Hausdorff Distance (HD95) between the detected end-diastolic and end-systolic phases with the motion-static segmentations.

For in vivo datasets, the dominant mode of 3D EF was compared with the available 2D cine EF as a clinical comparator. The 2D-derived EF was based on the first dominant sinus contraction from the sequential 2D stack. This means that several slices of different heart beats are used to build volumetric coverage, rather than the proposed method which obtains volumetric coverage from a single heartbeat. We therefore report the 3D mean EF as the primary in vivo finding and interpret cross-method differences mainly as measurement-target mismatch. Additionally, we plotted the mean EF per subject with standard deviation of the dominant mode distribution versus the clinical 2D EF together with a Bland-Altman plot.

# Results

## Phantom Study

The spatiotemporal plots in Figure 3 of the real-time 3D cine images illustrate the consistency of the motion patterns across multiple cardiac cycles. The modulations visible in each orthogonal view correspond to the periodic expansion and contraction of the phantom. These patterns remained stable across the entire acquisition, suggesting that the reconstructed motion fields capture the phantom's mechanical behaviour. The recovered motion and volume curves in Figure 6 also showed the expected dominant 1 Hz periodicity of the imposed piston setting as well as in the Frequency response plot in supplementary figure 9.

The phantom's EF obtained from the real-time reconstruction was in good agreement with the ground-truth EF determined from the motion-static acquisitions, with a mean ± standard deviation of 17.86 ± 0.33% compared with the ground-truth value of 17.3%. For the proposed segmentation propagation method, comparison with the ground-truth motion-static segmentations obtained a mean Dice score of 0.976 ± 0.003 and a mean HD95 of 3.88 ± 0.04 mm at end systole. At end diastole, the mean Dice score was 0.973 ± 0.001 and the mean HD95 was 3.89 ± 0.06 mm.

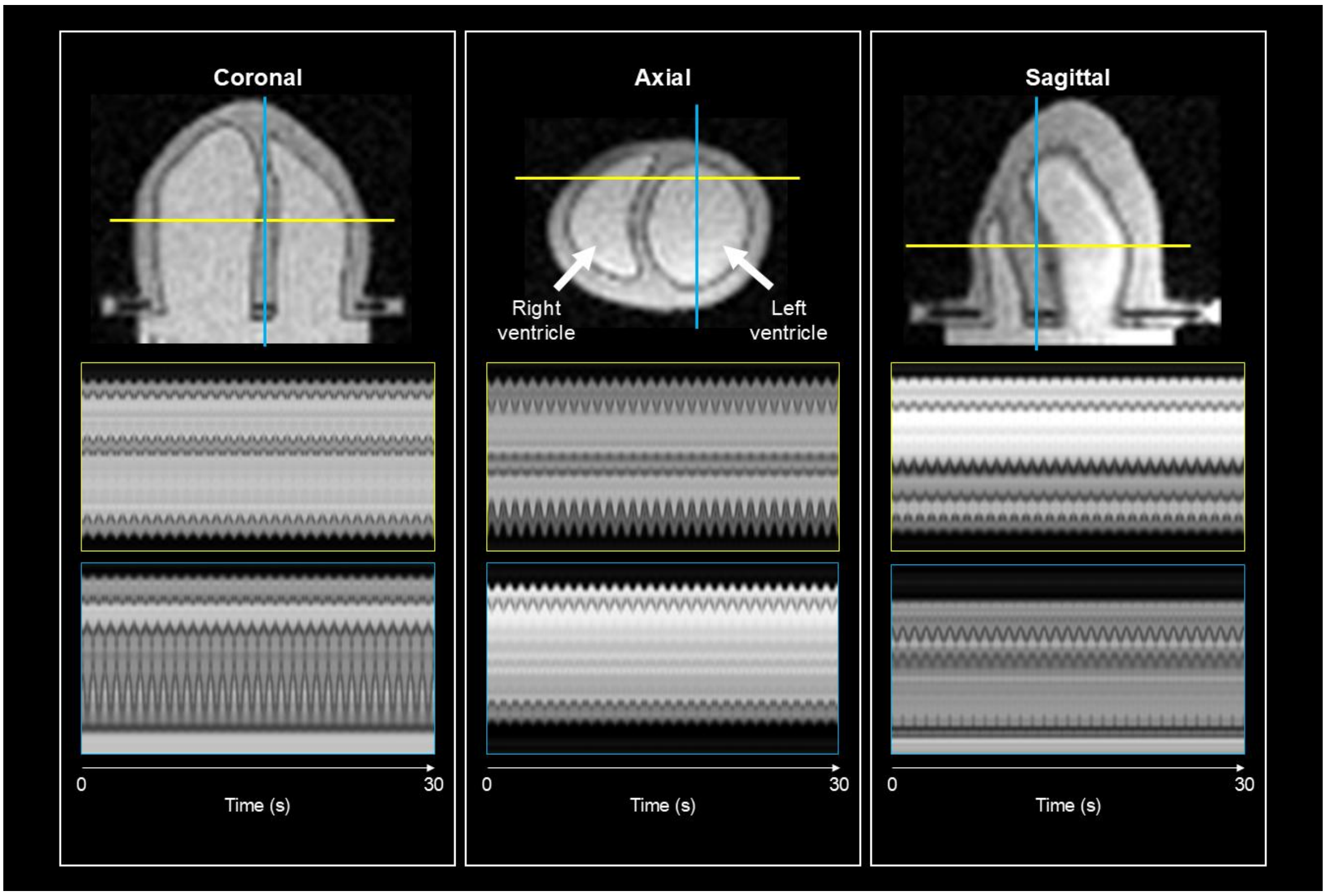

*Figure 3: Data shown in the figure is from the real-time phantom acquisition operating at 60 beats per minute. The figure shows the spatiotemporal plots of the real-time cines generated by warping the motion-corrected 3D reference image using the reconstructed motion fields from CMR-MOTUS at a nominal frame rate of ~20 Hz. For each orthogonal view, a red and green line indicates which 1D profiles in the spatiotemporal plots are defined. Below each view are the two spatiotemporal plots corresponding to coloured lines. Across all plots, modulations can be seen for each individual contraction. Note that a 30 second interval is presented here, whilst the entire reconstructed data is roughly 90 seconds long.*

## In Vivo Study

Figure 4 presents a representative example from a healthy volunteer, illustrating the quality of the reconstructed cine images. Since the acquired data is 3D, we reformatted the reconstructed cine sequences to standard CMR views including the 2-chamber, short-axis, and 4-chamber orientations. This flexibility demonstrates a key advantage of the 3D acquisition approach. The spatiotemporal plots shows the expected cardiorespiratory modulations. These examples are intended as qualitative demonstrations of continuous 3D real-time reconstruction rather than as formal validation against a GT cine.

Figure 5 shows a similar illustration, this time for a patient with PVC. The spatiotemporal plots show highly irregular modulations in the cardiac frequency. These irregularities correspond to the PVC episodes as seen by the temporal correspondence with the ECG signal and volume curve irregularities. The varying intervals between consecutive contractions and the differences in contraction patterns are visible in the spatiotemporal plots, highlighting the beat-to-beat variability. This variability would be averaged out in conventional binning-based reconstruction approaches, but the proposed real-time method preserves these clinically relevant dynamics.

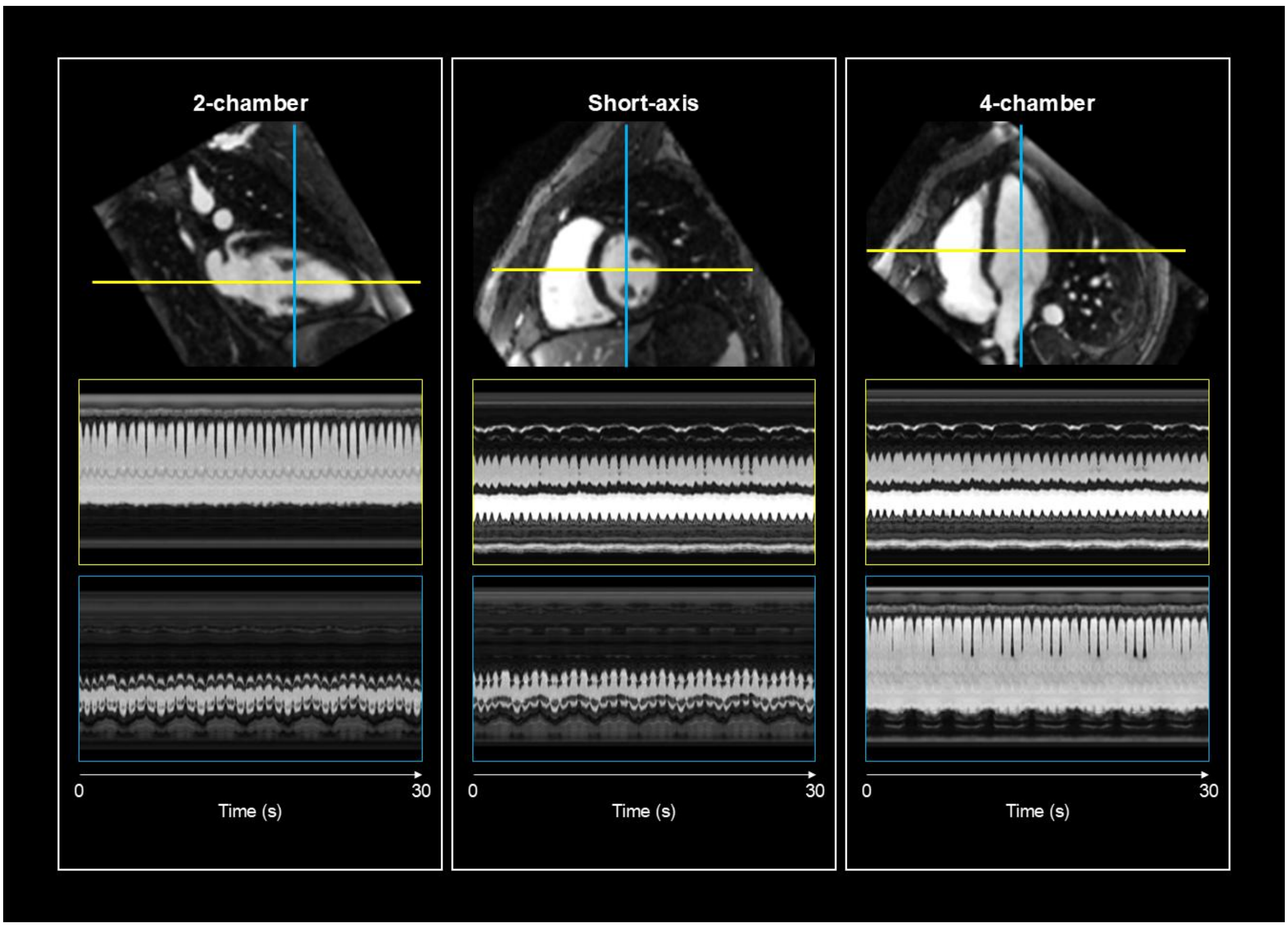

*Figure 4: Data shown in the figure is from a single healthy volunteer. The figure shows the spatiotemporal plots of the real-time cines generated by warping the motion-corrected 3D reference image using the reconstructed motion fields from CMR-MOTUS at a nominal frame rate of ~16 Hz. Since the data is 3D, we reformatted the cine to the standard CMR views (2-chamber, short-axis and 4-chamber). For each view, a red and green line indicates which locations the spatiotemporal plots are defined. Below each view are the two spatiotemporal plots corresponding to coloured lines. Across all plots, modulations can be seen for each individual heartbeat and the breathing cycles. Note that a 30 second interval is presented here, whilst the entire reconstructed data is roughly 2 minutes long.*

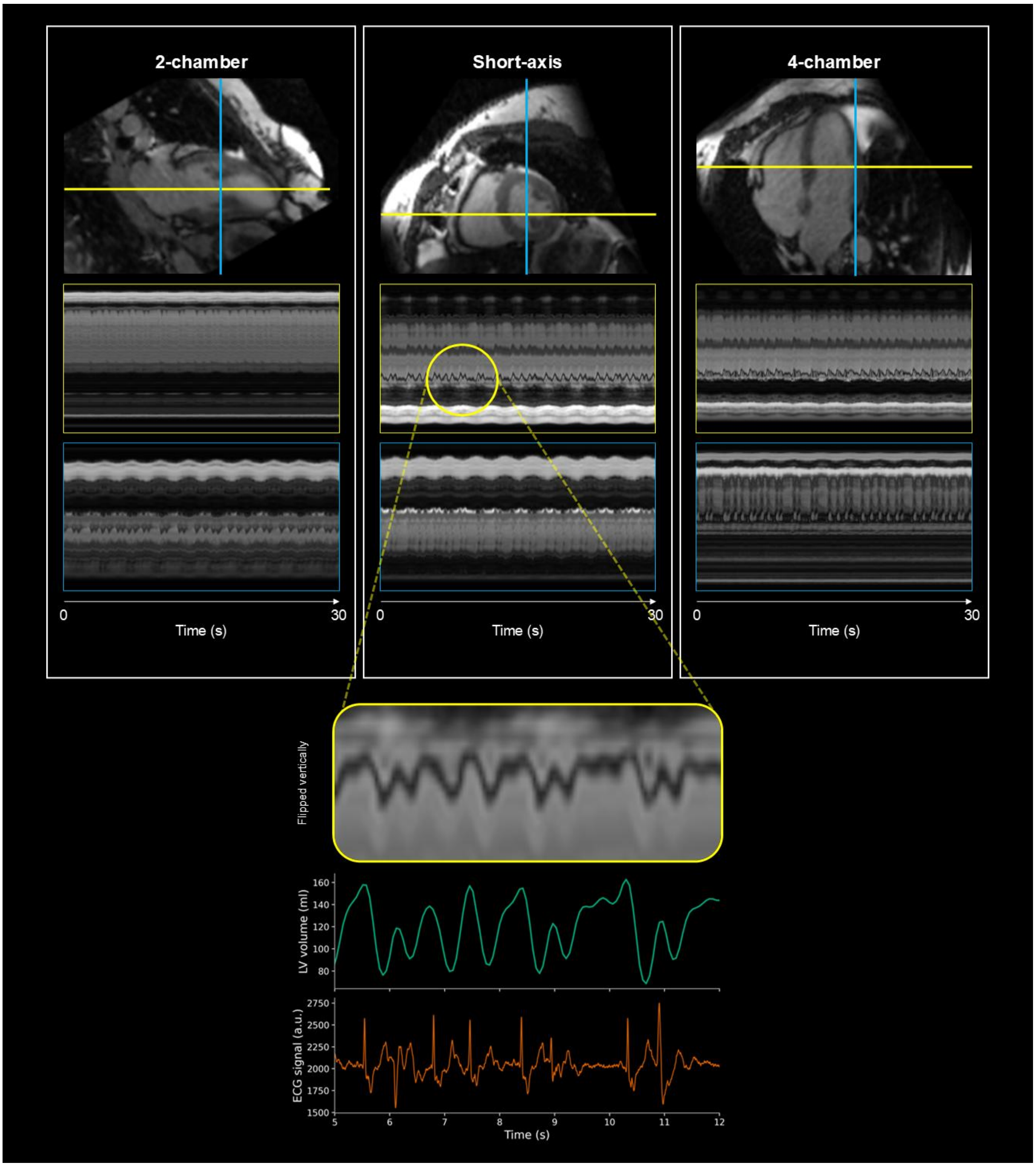

*Figure 5: Data shown in the figure is from a single subject with premature ventricular contraction (PVC). The figure shows the spatiotemporal plots of the real-time cines generated by warping the motion-corrected 3D reference image using the reconstructed motion fields from CMR-MOTUS at a nominal frame rate of ~17 Hz. Since the data is 3D, we reformatted the cine to the standard CMR views (2-chamber, short-axis and 4-chamber). For each view, a red and green line indicates which locations the spatiotemporal plots are defined. Below each view are the two spatiotemporal plots corresponding to coloured lines. Across all plots, modulations can be seen for each individual heartbeat and the breathing cycles. There is irregularity in the modulations due to several PVC episodes highlighting the significant beat-to-beat variability for these patient types. An example of these episodes is circled in yellow in a short-axis spatiotemporal plot. This area in the circle is enlarged below with the corresponding ECG signal and the proposed LV volume curve plotted. Note that a 30 second interval is presented here, whilst the entire reconstructed data is roughly 2 minutes long.*

The segmentation propagation and volume curve processing can be seen in Supplementary Figure 1-8 for a few selected subjects over a short time window. Figure 6 displays a representative volume curve for a healthy volunteer and a patient with PVC. The healthy volunteer shows regular periodic oscillations with consistent peaks and troughs, reflecting stable cardiac function with minimal beat-to-beat variability. In contrast, the patient with PVCs exhibits significant irregularities in the volume curve, with substantial variations in timing of individual cardiac cycles. Some contractions show reduced stroke volumes, corresponding to PVC episodes. Note that the simultaneously acquired ECG signal displayed below the volume curves supports the temporal correspondence between irregular volume patterns and PVC episodes.

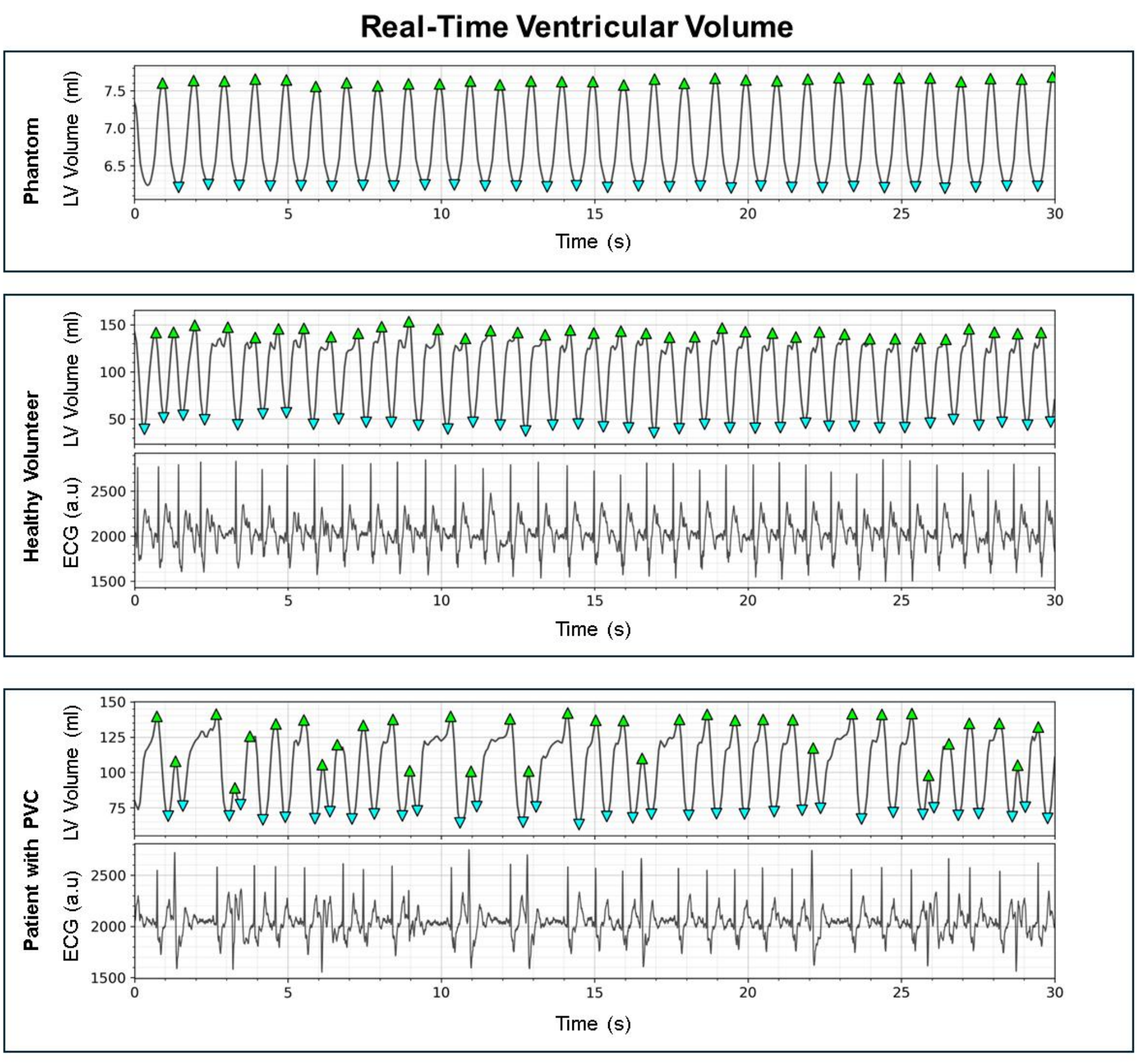

*Figure 6: Real-time ventricular volume plots for the phantom experiment, and a representative subject from the healthy volunteers and patients with premature ventricular contraction (PVC) groups. The temporal resolution for the phantom, healthy volunteer and patient with PVC datasets at a nominal frame rate of approximately 20 Hz, 16, Hz and 17 Hz respectively. Because the temporal coefficients were low pass filtered at 4 Hz, the effective temporal information bandwidth is lower than these nominal frame rates. Green and blue triangles represent the peak-trough pairs detected by the find_peaks algorithm in Scipy. Note the beat-to-beat variability in the in vivo cases; most severe in the patient case where there are PVC episodes present. For each peak-trough pair the ejection can be determined giving more understanding of how the ejection fraction varies time, but also how it varies during PVC episodes in the patient case. The plots showcase a 30 second interval of the entire acquisition which is roughly 2 minutes long. For the in vivo cases, the ECG signal that is acquired simultaneously with the acquisition is displayed under the volume plots. This shows that the irregular peak-trough pairs are linked to an irregular signal in the ECG suggesting that this is a PVC event.*

Figure 7 presents histograms of beat-to-beat EF distributions across the in vivo acquisitions. For healthy volunteers, the EF distributions are relatively narrow. The slight spread reflects the physiological beat-to-beat variability present even in healthy subjects. This temporal information, absent in conventional averaged acquisitions, provides a view of physiologic variability over the entire acquisition.

For patients with PVCs in Figure 7, the EF distributions show different characteristics. Subjects with PVC show bimodal distributed EF values, with the lower mode corresponding to EF during PVC episodes. These distributions reveal that individual PVC beats can have substantially lower ejection fractions than the patient's baseline cardiac function.

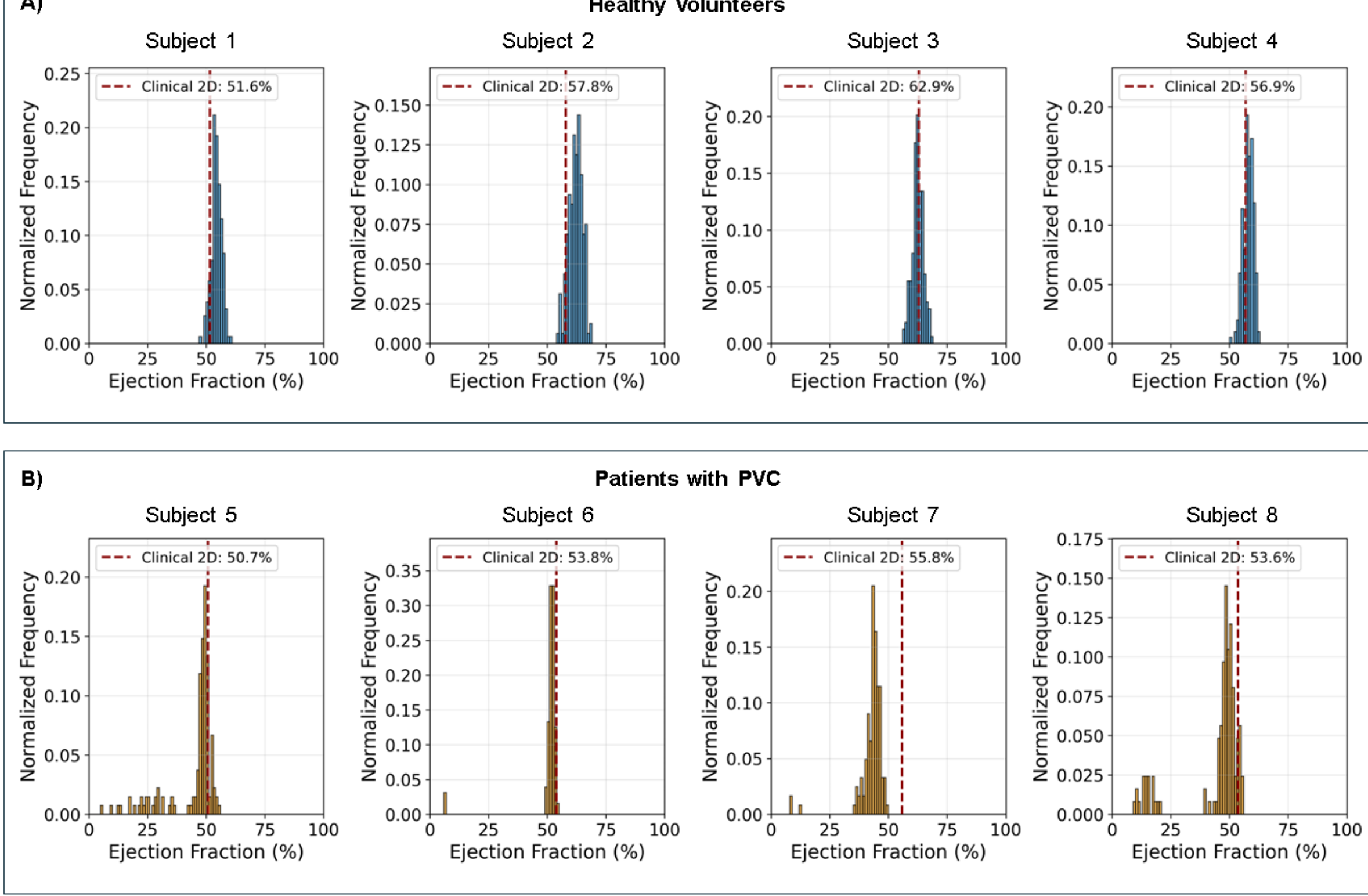

*Figure 7: Histogram of the frequency of ejection fractions (EF) determined in the entire ~2-minute in vivo acquisitions with selected healthy volunteers in A and patients with premature ventricular contraction (PVC) in B. The histogram bin width was set to 1%. A vertical dashed red line indicates the EF determined using the 2D acquisition. There is mixed agreement between the 2D and real-time 3D approaches. What is notable is that the real-time 3D EF offers statistics from the beat-to-beat variability. For the patients with PVC, the PVC episodes contribute the to secondary distribution of lower EF as seen particularly in subjects 5 and 8.*

Figure 8 compares the mean dominant EF from the proposed real-time 3D method against the GT and clinical reference values from standard 2D acquisitions. Phantom and healthy volunteer measurements showed smaller cross-method differences than the PVC group, where larger discrepancies are expected from the mismatch between a beat-to-beat summary and a sequential 2D stack that cannot represent the full irregular beat distribution. A regression of the in vivo data shows that the correlation is stronger in the healthy volunteers (slope 1.097, 95% CI 0.770-1.424; $R^2 = 0.882$) than in patients with PVC (slope 0.168, 95% CI -0.570-0.907; $R^2 = 0.033$). As an additional quantitative summary, Figure 9 shows a Bland-Altman analysis of the EF comparison. Across the 20 participants, the mean difference between the proposed method and 2D comparator was -3.05% (SD: 4.87%), with 95% limits of agreement from -12.60% to 6.51%.

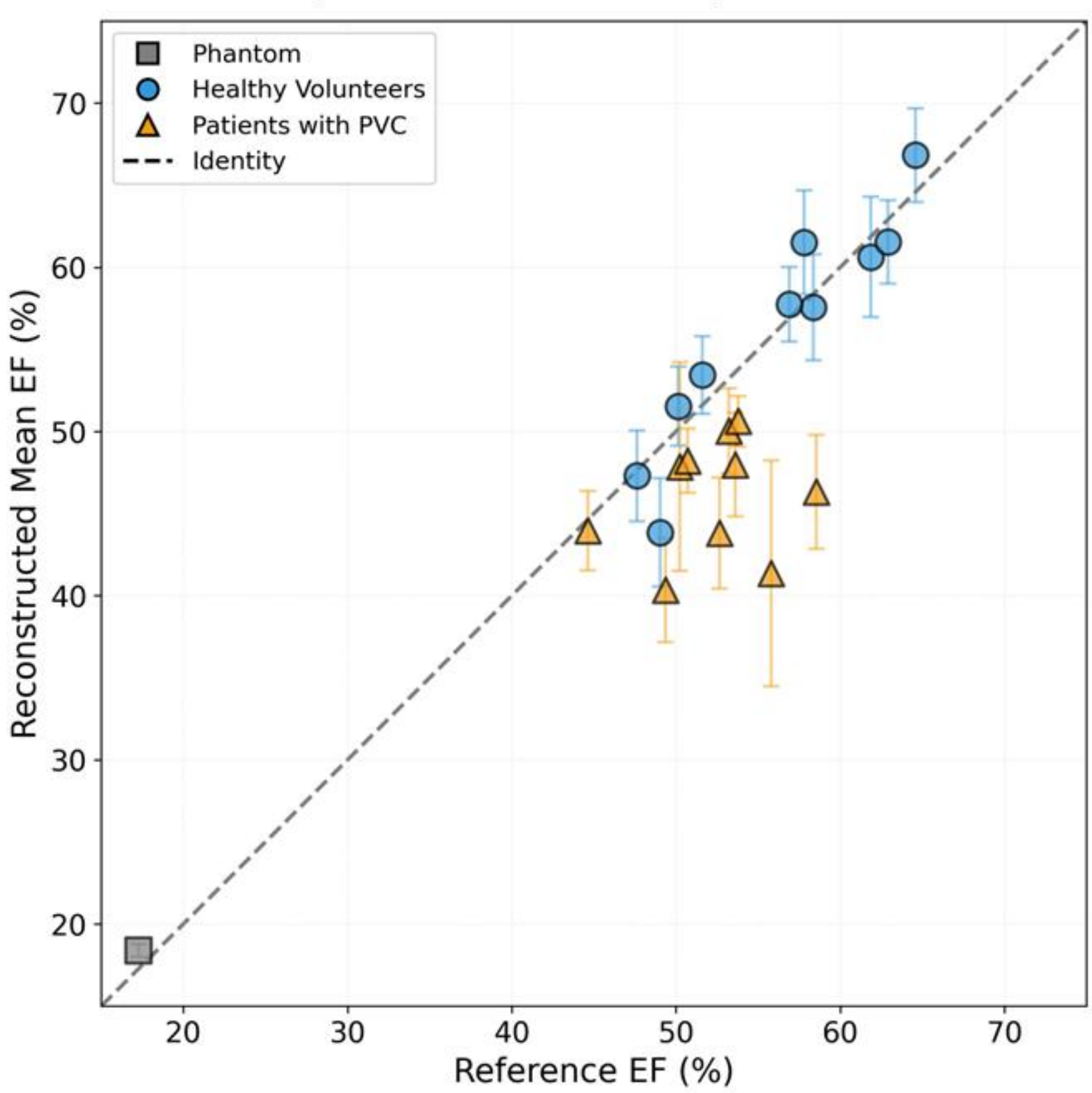

*Figure 8: Comparison of the mean dominant ejection fractions (EF) determined by the proposed continuous 3D volume curve and the standard 2D clinical method in the human subjects. For the phantom study, the reference EF was determined by the motion static acquisitions at different piston amplitudes. Error bars indicate the standard deviation of the distribution of EFs determined over the entire acquisition. There is agreement in the phantom study and most of the healthy volunteer cases. However, this is different for the patients with premature ventricular contraction (PVC) cases.*

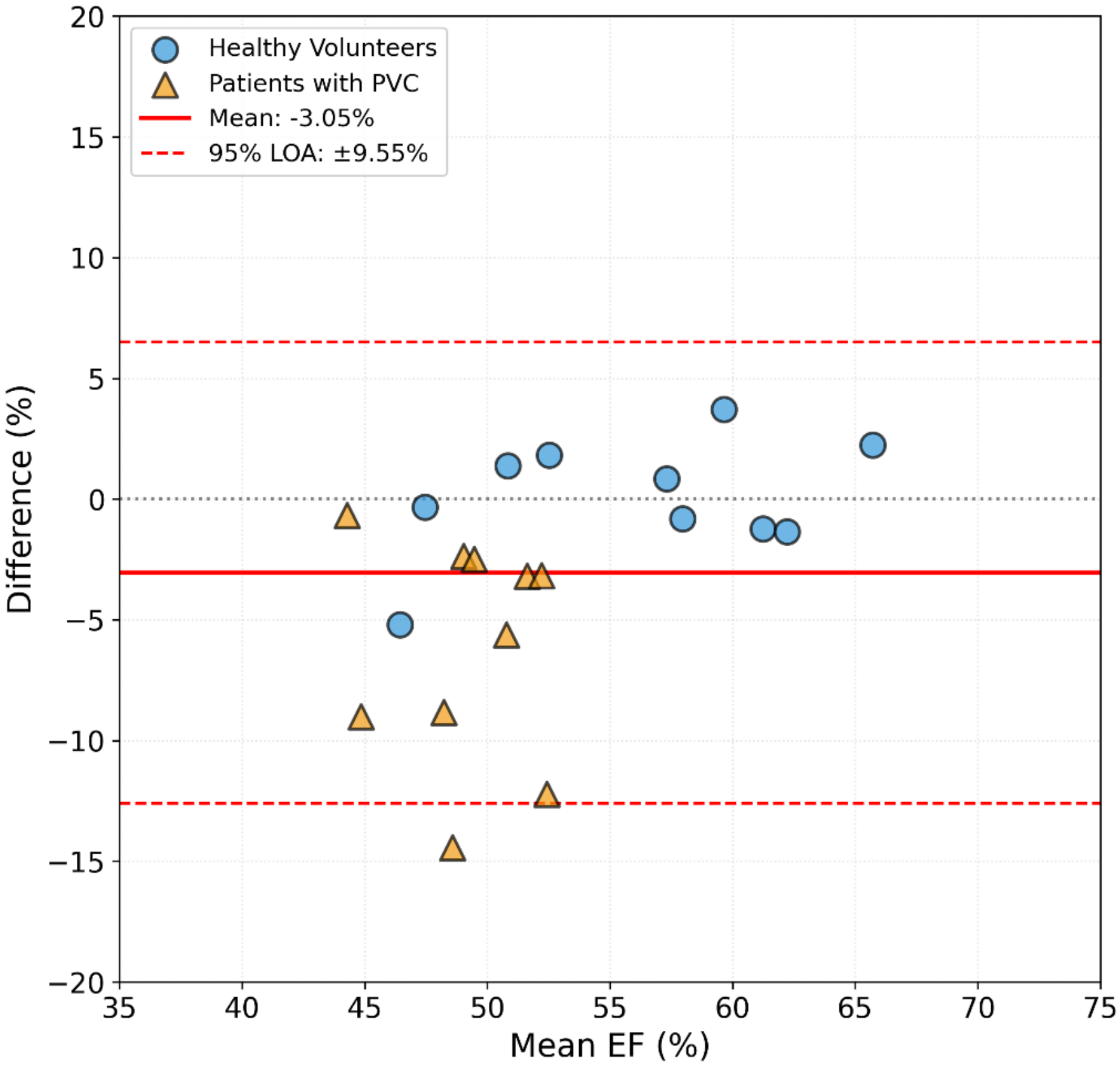

*Figure 9: Bland-Altman plot comparing the ejection fraction (EF) determined by the proposed reconstructed mean EF and the standard clinical 2D real-time cine method in the 20 in vivo subjects. The x-axis shows the mean EF of the two methods for each subject, and the y-axis shows the difference between the proposed reconstructed mean EF and the 2D clinical method. The solid horizontal line indicates the mean bias, and the dashed lines indicate the 95% limits of agreement (LOA). For patients with PVC and bimodal reconstructed EF distributions, the dominant high-EF mode was used for comparison to the 2D reference.*

# Discussion

This study demonstrates the feasibility of a 3D real-time motion field reconstruction in PVC patients using a free-running CMR protocol, enabling automated beat-to-beat volumetric quantification through segmentation propagation. In contrast to other real-time 3D MRI methods, the main contribution of our proposed approach is the explicit separation of physiological motion into interpretable motion fields and reconstruction of a single motion-corrected reference image enabling downstream analysis from the reconstructed motion fields. The present study should be interpreted as an early technical feasibility study rather than as formal clinical validation..

The beat-to-beat volumetric assessment may have important clinical applications. First, patients with PVCs may have preserved EF on standard gated CMR but reduced mean function when all beats are considered. Second, therapeutic monitoring after treatment procedures could assess treatment's effectiveness not only in terms of PVC frequency reduction but also by improvement in EF distribution characteristics and reduction in beat-to-beat variability. Third, the statistical distribution of EF across cycles could provide insight into cardiac functional compensatory capacity that one-beat measurements cannot capture. These metrics may prove valuable for guiding treatment decisions and predicting clinical outcomes, though prospective clinical validation studies are needed. On the other hand, the motion fields provide a foundation for comprehensive deformation analysis. Future work could derive dyssynchrony, strain, strain rate, and regional wall motion metrics from the motion fields to provide complete myocardial functional characterization.

In our phantom experiment, the small standard deviation across detected cycles demonstrates the consistency of the reconstruction method. We observed a slight beat-to-beat variation in the reconstructed EF of the phantom, which can be attributed to the different mechanical conditions between the two measurement approaches. In the real-time case, the piston operated continuously, potentially introducing dynamic elastic responses in the silicone material during the expansion and contraction phases. In contrast, the static reference acquisitions were performed at static displacement positions where such mechanical effects were absent. This may contribute to subtle variations in the observed volumes. However, because the phantom does not reproduce arrhythmic beat morphology, inflow behaviour, contrast dynamics, or the exact in vivo LV analysis pipeline, this experiment should be interpreted as a simplified technical validation of segmentation propagation and EF recovery rather than as thorough validation of the expected clinical scenario.

Agreement between the proposed mean EF and the reference EF  was the strongest in healthy volunteers and the phantom experiments. However, for PVC patients, a systematic underestimation is observed in reconstructed mean EF compared to the 2D reference as seen in Figure 8 and Figure 9. This apparent disagreement reflects a fundamental difference in what each method measures. The standard 2D acquisition uses real-time 2D slices acquired sequentially over multiple heartbeats to form one EF estimate based on one sinus contraction. . In contrast, the proposed method captures all cardiac events, resulting in a distribution of EFs that when averaged could be accounting for reduced function.

The underlying framework is not inherently limited to cardiac applications. The same free-running acquisition and CMR-MOTUS reconstruction could in principle be extended to other organs and body regions where physiological motion presents challenges, such as abdominal or head, where speech, respiratory and cardiac motion similarly complicate conventional acquisitions.

## Limitations

This feasibility study included only 10 healthy volunteers and 10 PVC patients. While this cohort was sufficient to demonstrate technical feasibility of the method and to reveal clinically relevant beat-to-beat variability, it limits the generalizability of the findings. Larger cohort studies are needed to validate reproducibility and assess the prognostic value of beat-to-beat functional heterogeneity.

The phantom assesses segmentation propagation and EF recovery only in a simplified periodic setting and does not reproduce arrhythmic beat morphology, inflow behaviour, or temporal contrast dynamics. In addition, no matched 2D real-time phantom acquisition was available for trace-to-trace comparison. Phantom agreement should therefore be interpreted as technical support for this propagation and EF recovery component rather than as validation of the full in vivo pipeline.

All in vivo demonstrations in this study were contrast enhanced, so performance without contrast enhancement remains untested. Based on the present data, sufficient blood-myocardium contrast appears important for reliable motion estimation and segmentation propagation. Whether the framework is viable under native contrast conditions remains to be determined.

The datasets used in this manuscript do not contain dynamic contrast enhancement. Hence, a time-independent reference image was reconstruction. This static-reference formulation cannot explicitly represent non-deformation and non-conversative signal changes such as inflow or phase decoherence, and the 4 Hz temporal filtering may attenuate higher-frequency dynamics. Larger controlled studies, including phantoms, are required to measure the proposed method’s sensitivity to these effects.

Validation of segmentation propagation in in vivo data remains indirect. Because real-time 3D cine MRI is challenging to reconstruct at a 20 Hz frame rate, the robustness of propagated LV segmentations under irregular human physiology is not yet fully quantified.

Because the sequential 2D short-axis stack and the proposed 3D beat-to-beat reconstruction do not represent the same EF quantity, Figure 8 and Figure 9 should be interpreted only as quantitative support and not as evidence of accuracy, interchangeability, or clinical equivalence.

Reconstruction remained computationally expensive and offline in the present implementation. Additionally, the regularization hyperparameter $\lambda$ was empirically tuned for each dataset. These limitations may limit immediate clinical deployment even if methodological feasibility is established.

The phantom, healthy volunteers and PVC patients were scanned on different MRI systems with different field strengths, sequences, and contrast agents, which may introduce confounding factors when comparing results between groups. This heterogeneity reflects clinical reality where scan protocols are tailored to the clinical question and available hardware, but future controlled studies should aim for more homogeneous acquisition protocols to isolate the effect of the reconstruction method from acquisition-related variability.

# Conclusion

This study establishes real-time 3D motion field reconstruction as a potentially useful approach for cardiac functional assessment in arrhythmic patients. By explicitly extracting physiological motion into interpretable motion fields, the framework enables single-segmentation propagation across all cardiac phases, making beat-to-beat volumetric quantification practical and revealing functional heterogeneity that conventional measurements may obscure. However, the present results should be interpreted as early feasibility evidence rather than as clinical validation, and larger controlled studies are needed before claims of replacement of conventional 2D cine analysis.

# Supplementary Information

subject_1.mp4

*Supplementary Figure 1: Data in figure is from subject 1. The figure shows a video of the CMR-MOTUS reconstruction, which is a reference image warped with the reconstruction real-time motion fields that are overlaid as green vectors. The segmentation of the left ventricle blood pool is propagated with these motion fields to obtain the real-time ventricular volume .*

subject_2.mp4

*Supplementary Figure 2: Data in figure is from subject 2. The figure shows a video of the CMR-MOTUS reconstruction, which is a reference image warped with the reconstruction real-time motion fields that are overlaid as green vectors. The segmentation of the left ventricle blood pool is propagated with these motion fields to obtain the real-time ventricular volume .*

subject_3.mp4

*Supplementary Figure 3: Data in figure is from subject 3. The figure shows a video of the CMR-MOTUS reconstruction, which is a reference image warped with the reconstruction real-time motion fields that are overlaid as green vectors. The segmentation of the left ventricle blood pool is propagated with these motion fields to obtain the real-time ventricular volume.*

subject_4.mp4

*Supplementary Figure 4: Data in figure is from subject 4. The figure shows a video of the CMR-MOTUS reconstruction, which is a reference image warped with the reconstruction real-time motion fields that are overlaid as green vectors. The segmentation of the left ventricle blood pool is propagated with these motion fields to obtain the real-time ventricular volume.*

subject_5.mp4

*Supplementary Figure 5: Data in figure is from subject 5 with premature ventricular contractions (PVC). The figure shows a video of the CMR-MOTUS reconstruction, which is a reference image warped with the reconstruction real-time motion fields that are overlaid as green vectors. The segmentation of the left ventricle blood pool is propagated with these motion fields to obtain the real-time ventricular volume .*

subject_6.mp4

*Supplementary Figure 6: Data in figure is from subject 6 with premature ventricular contractions (PVC). The figure shows a video of the CMR-MOTUS reconstruction, which is a reference image warped with the reconstruction real-time motion fields that are overlaid as green vectors. The segmentation of the left ventricle blood pool is propagated with these motion fields to obtain the real-time ventricular volume.*

subject_7.mp4

*Supplementary Figure 7: Data in figure is from subject 7 with premature ventricular contractions (PVC). The figure shows a video of the CMR-MOTUS reconstruction, which is a reference image warped with the reconstruction real-time motion fields that are overlaid as green vectors. The segmentation of the left ventricle blood pool is propagated with these motion fields to obtain the real-time ventricular volume.*

subject_8.mp4

*Supplementary Figure 8: Data in figure is from subject 8 with premature ventricular contractions (PVC). The figure shows a video of the CMR-MOTUS reconstruction, which is a reference image warped with the reconstruction real-time motion fields that are overlaid as green vectors. The segmentation of the left ventricle blood pool is propagated with these motion fields to obtain the real-time ventricular volume.*

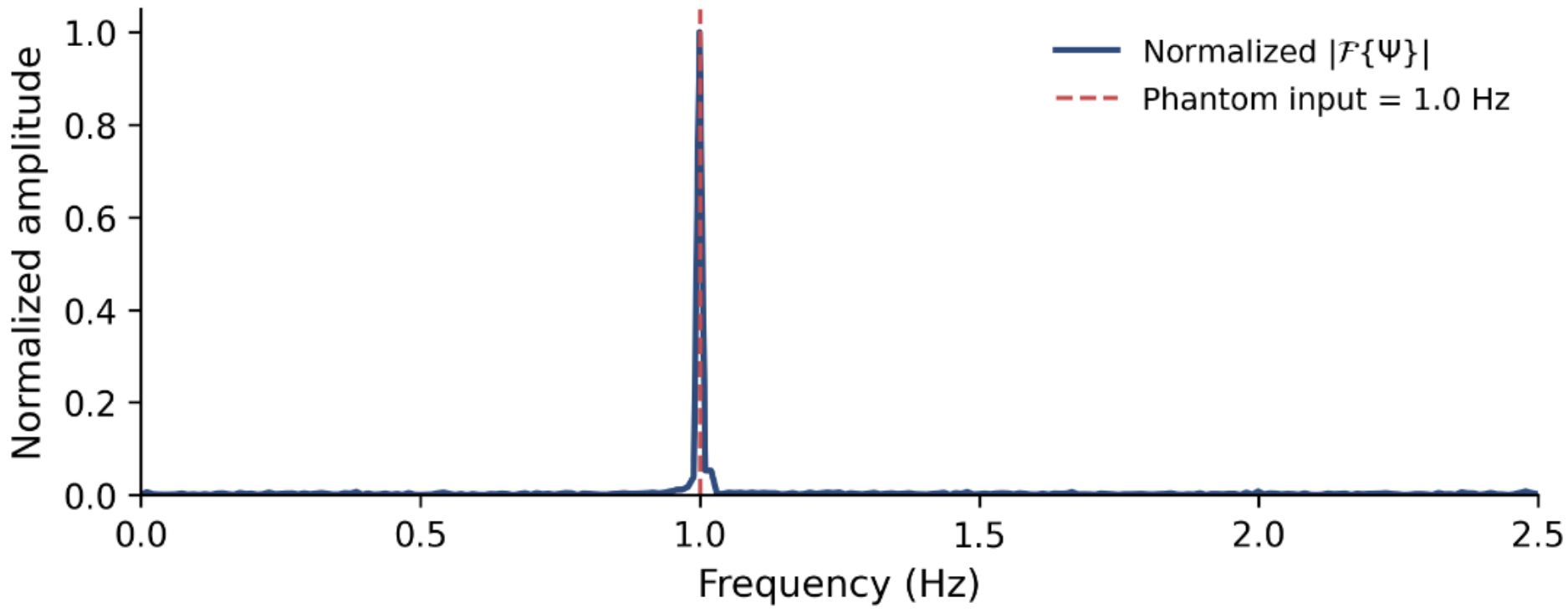

*Supplementary Figure 9: Plot of the frequency response of the $\Psi$ component (temporal scaling of motion fields) of the phantom experiment. The phantom was set to oscillate in a sinusoidal fashion at 1 Hz temporal frequency marked with a red dashed line in the plot. There is a clear frequency correspondence between the $\Psi$ and the phantom setting.*

# Data Availability

The computer code and the mechanical phantom data are available upon a reasonable request to the corresponding author.

# Funding

This research is funded by the Netherlands Organisation for Scientific Research (NWO), domain Applied and Engineering Sciences, Grant number 19003, and by NIH grants R01-EB029957, R01-HL151697, and R01-HL148103.

# Author Contributions

**Thomas E. Olausson:** Conceptualization, Methodology, Software, Validation, Formal analysis, Investigation, Data curation, Writing - original draft, Visualization. **Maarten L. Terpstra:** Conceptualization, Methodology, Supervision, Writing - review & editing. **Rizwan Ahmad:** Resources, Data curation, Writing - review & editing. **Edwin Versteeg:** Methodology, Software, Investigation, Writing - review & editing. **Casper Beijst:** Supervision, Writing - review & editing. **Yuchi Han:** Resources, Data curation, Writing - review & editing. **Marco Guglielmo:** Validation, Writing - review & editing. **Birgitta K. Velthuis:** Validation, Writing - review & editing. **Cornelis van den**

**Berg:** Conceptualization, Methodology, Supervision, Writing - review & editing. **Alessandro Sbrizzi:** Conceptualization, Methodology, Supervision, Writing - review & editing.

## Ethics Approval and Consent

For the human subject data, approval was granted by the Institutional Review Board (IRB) at The Ohio State University (2021X0110, 2020H0402 and 2019H0076). Informed consent to participate in the study and publish results was obtained from all individual participants.

## Acknowledgements

The authors would like to thank Syed Murtaza Arshad for the fruitful discussions.